\documentclass[trackchanges,twocolumn]{aastex7}

\begin{document}

\title{Direct measurement of $^{59}$Cu($p$,$\alpha$)$^{56}$Ni precludes a strong NiCu cycle in Type-I X-ray bursts}

\author[]{N. Bhathi}
\affiliation{Astronomy and Physics Department, Saint Mary’s University, Halifax, Nova Scotia B3H 3C3, Canada}
\affiliation{TRIUMF, Vancouver, BC V6T2A3, Canada}
\email[]{}

\correspondingauthor{J. S. Randhawa}
\author[0000-0001-6860-3754]{J. S. Randhawa}
\affiliation{Department of Physics and Astronomy, Mississippi State University, Mississippi 39762, USA}
\email[show]{jsr512@msstate.edu}

\author{R. Kanungo$^*$}
\affiliation{TRIUMF, Vancouver, BC V6T2A3, Canada}
\affiliation{Astronomy and Physics Department, Saint Mary’s University, Halifax, Nova Scotia B3H 3C3, Canada}
\email[show]{$^*$Email : ritu@triumf.ca}

\author{J. Refsgaard}
\affiliation{Astronomy and Physics Department, Saint Mary’s University, Halifax, Nova Scotia B3H 3C3, Canada}
\affiliation{TRIUMF, Vancouver, BC V6T2A3, Canada}
\email[]{}
\author{M. Alcorta}
\affiliation{TRIUMF, Vancouver, BC V6T2A3, Canada}
\email[]{}

\author[]{T. Ahn}
\affiliation{Department of Physics, University of Notre Dame, Notre Dame, IN 46556}
\email[]{}
\author{C. Andreoiu}
\affiliation{Department of Chemistry, Simon Fraser University, Burnaby, British Columbia V5A 1S6, Canada}
\email[]{}

\author[]{D. Bardayan}
\affiliation{Department of Physics, University of Notre Dame, Notre Dame, IN 46556}
\email[]{}

\author{S. S. Bhattacharjee}
\affiliation{Astronomy and Physics Department, Saint Mary’s University,
 Halifax, Nova Scotia B3H 3C3, Canada}%
\affiliation{TRIUMF, Vancouver, BC V6T2A3, Canada}
\email[]{}

\author{B. Davids}
\affiliation{TRIUMF, Vancouver, BC V6T2A3, Canada}
\affiliation{Department of Physics, Simon Fraser University, Burnaby, British Columbia V5A 1S6, Canada}
\email[]{}

\author{G. Christian}
\affiliation{Astronomy and Physics Department, Saint Mary’s University,
 Halifax, Nova Scotia B3H 3C3, Canada}
 \email[]{}

\author{A. A. Chen }
\affiliation{Department of Physics and Astronomy, McMaster University, Hamilton, Ontario L8S 4M1, Canada}
\email[]{}

\author{R. Coleman}
\affiliation{Department of Physics, University of Guelph, Guelph, Ontario N1G 2W1, Canada}
\email[]{}

\author{P. E. Garrett}
\affiliation{Department of Physics, University of Guelph, Guelph, Ontario N1G 2W1, Canada}
\affiliation{Department of Physics, University of the Western Cape, P/B X17, Bellville, ZA-7535 South Africa}
\email[]{}

\author{G. F. Grinyer}
\affiliation{Department of Physics, University of Regina, Regina, SK S4S 0A2, Canada}

\email[]{}
\author{E. Gyabeng Fuakye}
\affiliation{Department of Physics, University of Regina, Regina, SK S4S 0A2, Canada}

\email[]{}

\author{G. Hackman}
\affiliation{TRIUMF, Vancouver, BC V6T2A3, Canada}
\email[]{}

\author{R. Jain}
\affiliation{Facility for Rare Isotope Beams, Michigan, 48824, U.S.}
\email[]{}

\author{K. Kapoor}
\affiliation{Department of Physics, University of Regina, Regina, SK S4S 0A2, Canada}

\email[]{}

\author{R. Kr\"ucken}
\affiliation{TRIUMF, Vancouver, BC V6T2A3, Canada}
\affiliation{Department of Physics and Astronomy, University of British Columbia, Vancouver, BC V6T 1Z1, Canada}
\email[]{}

\author{A. Laffoley}
\affiliation{Department of Physics, University of Guelph, Guelph, Ontario N1G 2W1, Canada}
\email[]{}

\author{A. Lennarz}
\affiliation{TRIUMF, Vancouver, BC V6T2A3, Canada}
\affiliation{Department of Physics and Astronomy, McMaster University, Hamilton, Ontario L8S 4M1, Canada}
\email[]{}

\author{J. Liang}
\affiliation{Department of Physics and Astronomy, McMaster University, Hamilton, Ontario L8S 4M1, Canada}
\email[]{}

\author{Z. Meisel}
\affiliation{Institute of Nuclear and Particle Physics, Department of Physics \& Astronomy, Ohio University, Athens, OH 45701, USA}
\email[]{}
 
\author{A. Psaltis}
\affiliation{Department of Physics and Astronomy, McMaster University, Hamilton, Ontario L8S 4M1, Canada}
\email[]{}

\author{A. Radich}
\affiliation{Department of Physics, University of Guelph, Guelph, Ontario N1G 2W1, Canada}
\email[]{}
\author{M. Rocchini}
\affiliation{Department of Physics, University of Guelph, Guelph, Ontario N1G 2W1, Canada}
\email[]{}
\author{J.S. Rojo}
\affiliation{TRIUMF, Vancouver, BC V6T2A3, Canada}
\email[]{}

\author{N. Saei} 
\affiliation{Department of Physics, University of Regina, Regina, SK S4S 0A2, Canada}
\email[]{}

\author{M. Saxena}
\affiliation{Institute of Nuclear and Particle Physics, Department of Physics \& Astronomy, Ohio University, Athens, OH 45701, USA}
\email[]{}

\author{M. Singh}
\affiliation{Astronomy and Physics Department, Saint Mary’s University,
 Halifax, Nova Scotia B3H 3C3, Canada}
 \email[]{}

\author{C. E. Svensson}
\affiliation{Department of Physics, University of Guelph, Guelph, Ontario N1G 2W1, Canada}
\affiliation{TRIUMF, Vancouver, BC V6T2A3, Canada}
\email[]{}

\author{P. Subramaniam}
\affiliation{Astronomy and Physics Department, Saint Mary’s University,
 Halifax, Nova Scotia B3H 3C3, Canada}
\email[]{}

\author{A. Talebitaher}
\affiliation{Department of Physics, University of Regina, Regina, SK S4S 0A2, Canada}
 \email[]{}

\author{S. Upadhyayula}
\affiliation{TRIUMF, Vancouver, BC V6T2A3, Canada}
\email[]{}

\author{C. Waterfield}
\affiliation{Astronomy and Physics Department, Saint Mary’s University, Halifax, Nova Scotia B3H 3C3, Canada}
\email[]{}

 \author{J. Williams}
\affiliation{TRIUMF, Vancouver, BC V6T2A3, Canada}
\email[]{} 

\author{M. Williams}
\affiliation{TRIUMF, Vancouver, BC V6T2A3, Canada}
\email[]{}

\author{M. A. Zubair}
\affiliation{Department of Physics and Astronomy, Mississippi State University, Mississippi 39762, USA}
\email[]{}

\begin{abstract}
Model-observation comparisons of type-I X-ray bursts (XRBs) can reveal the properties of accreting neutron star systems, including the neutron star compactness. XRBs are powered by nuclear burning and a handful of reactions have been shown to impact the model results. Reactions in the NiCu cycles, featuring a competition between $^{59}$Cu($p$,$\gamma$)$^{60}$Zn and $^{59}$Cu($p$,$\alpha$)$^{56}$Ni, have been shown  to be among the most important reactions as they are a critical checkpoint in $rp$-process flow and significantly impact the light curves and burst ashes. We report a direct measurement of  $^{59}$Cu($p$,$\alpha$)$^{56}$Ni bringing stringent constraints on this reaction rate. New results rule out a strong NiCu cycle in XRBs, with a negligible degree of recycling, $\leq$5\% up to 1.5 GK. The new reaction rate, when varied within new  uncertainty limits, shows no impact on one-zone XRB model light-curves tailored for clocked-burster $\tt{GS 1826-24}$, hence removing an important nuclear physics uncertainty in the model-observation comparison. 

\end{abstract}



\section{Motivation} 
Type-I X-ray bursts (XRBs) are thermonuclear explosions on the surface of accreting neutron stars powered by nuclear burning \citep{SCHATZ2006}.  Advances in XRB observations with space-based telescopes and advanced computer models have opened up a new portal to constrain the neutron star mass-radius relation and other underlying physics through comparison between observations and models \citep{Galloway_2004, Heger_2007,Meisel_2019,Galloway_2004}. In particular, the X-ray burster $\tt{GS 1826–24}$ shows extremely regular Type I X-ray bursts whose energetics and recurrence times agree well with thermonuclear ignition models, and has been frequently used for model-observation comparisons to infer the astrophysical conditions \citep{Heger_2007, Galloway_2004} and neutron star compactness \citep{Meisel_2019,Randhawa_22Mg}. In XRBs, the accreted material is burned via nuclear reactions, dominantly through the $\alpha p$-process  and $rp$-process, where the former is  a reaction sequence of ($\alpha,p$) reaction followed by (p,$\gamma$) which takes the reaction flow generally up to the calcium region and  later the rapid-proton capture proceeds beyond the calcium region to all the way to Sn \citep{Meisel_review, Schatz-Ani}. The energy generated during this nuclear burning shapes the XRB light curves, and hence reliable nuclear input is required to validate the assumptions of astrophysical models through model versus observation comparisons, and to constrain the underlying neutron star mass-radius relation \citep{Randhawa_22Mg, Meisel_2019}. Accurate nuclear physics input plays an equally important role in predicting the burst ashes, which become part of the neutron star crust and alter the composition of the neutron star crust \citep{Jain_2023}. Reaction rate sensitivity studies of XRBs \citep{Cyburt_2016, Meisel_2019} have revealed that only a handful of key nuclear reactions have a significant impact on the overall energy generation and nucleosynthesis in such environments. Out of these crucial reactions, the  Nickel-Copper (NiCu) cycle reactions are shown to be the most important. Variation in these lead to significant impacts on XRB light curves  and ashes. This cycle comprises a closed loop of nuclear reactions i.e., $^{56}$Ni($p$,$\gamma$)$^{57}$Cu($p$,$\gamma$)$^{58}$Zn($\beta^{+}\nu$)$^{58}$Cu($p$,$\gamma$)$^{59}$Zn($\beta^{+}\nu$)$^{59}$Cu($p$,\\$\alpha$)$^{56}$Ni (or $^{59}$Cu($p$,$\gamma$)$^{60}$Zn), where the competition between the $^{59}$Cu($p$,$\gamma$)$^{60}$Zn and $^{59}$Cu(p,$\alpha$)$^{56}$Ni reactions defines the cycling strength in the NiCu cycle. If $^{59}$Cu($p$,$\alpha$)$^{56}$Ni, is faster than $^{59}$Cu(p,$\gamma$)$^{60}$Zn, it takes the reaction flow back to the beginning of NiCu cycle i.e., to $^{56}$Ni. For the $rp$-process to proceed beyond this mass region, the $^{59}$Cu(p,$\gamma$)$^{60}$Zn reaction must dominate over $^{59}$Cu($p$,$\alpha$)$^{56}$Ni. The NiCu cycle thus acts as a critical checkpoint in the $rp$-process flow. Various studies have shown that the competition between the $^{59}$Cu($p$,$\alpha$)$^{56}$Ni and $^{59}$Cu($p$,$\gamma$)$^{60}$Zn reactions has a significant impact on the XRB model light curves and burst ashes, as it regulates the NiCu cycle \citep{Cyburt_2016, Meisel_2019,Kim_2022}, hence are among the most important reactions. Currently, both the ($p$,$\alpha$) and ($p$,$\gamma$) reaction rates in the REACLIB reaction rate library are derived from statistical model based NON-SMOKER predictions \citep{RAUSCHER2001}. Even though previous reaction rate sensitivity studies have used  an uncertainty of $10^{4}$ (i.e. factor 100 up and down) in both the ($p$,$\gamma$) and (p,$\alpha$) reaction rates, even a factor of 10 change  could lead to either a strong NiCu cycle (for ($p$,$\alpha$) UP and ($p$,$\gamma$) DOWN), or a negligible cycling (for $p$,$\alpha$) DOWN and ($p$,$\gamma$) UP). Therefore, within the current reaction rate uncertainties, either a strong or a weak NiCu cycle is possible.\\
\begin{figure}
    \centering
    \includegraphics[width=0.9\linewidth]{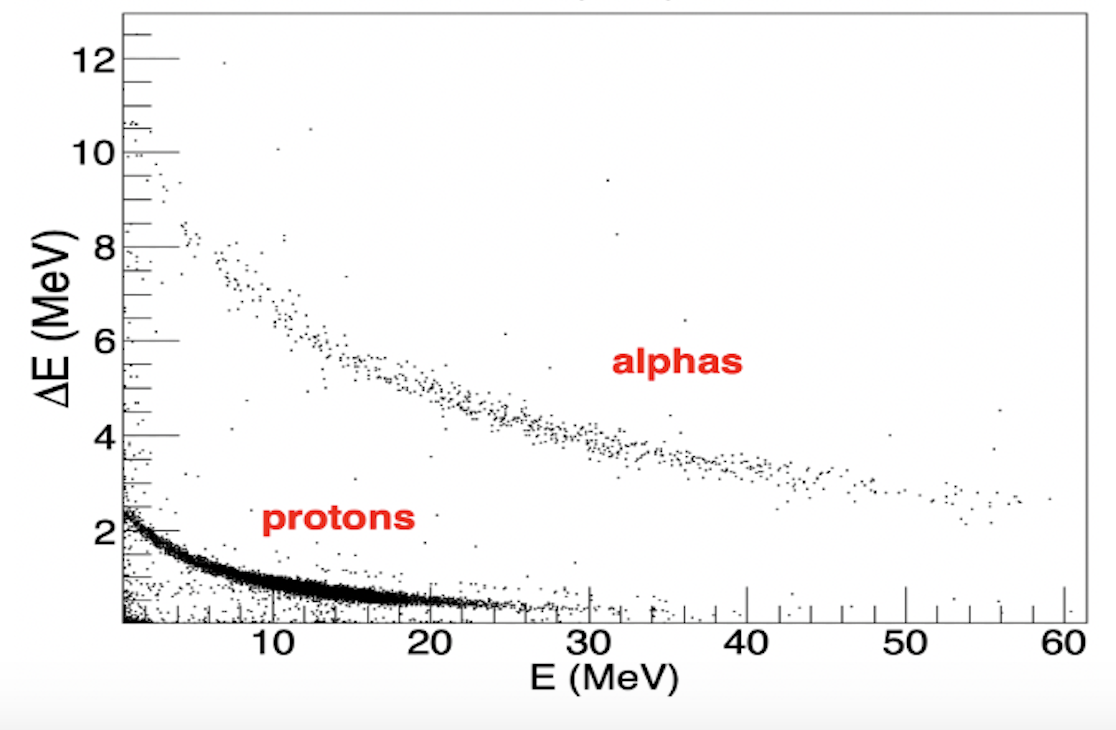}
    \caption{Particle Identification plot, using $\Delta$E-E detectors, at E$_{c.m.}$ 4.68$\pm$0.25 MeV}
    \label{fig1}
\end{figure}
To understand the strength of NiCu cycle in XRBs, \cite{Kim_2022} estimated the  reaction rate  from existing nuclear structure data (which is sparse) on  the compound nucleus $^{60}$Zn. The study of \cite{Kim_2022} predicted a very different (lower) reaction rate compared to NON-SMOKER predictions for the $^{59}$Cu(p,$\alpha$)$^{56}$Ni reaction, whereas their $^{59}$Cu($p$,$\gamma$)$^{60}$Zn estimate agreed with this model prediction (within uncertainty limits), hence predicting a weaker NiCu cycle. More recently, an indirect measurement by \cite{OShea25} reported new stringent constraints on the $^{59}$Cu(p,$\gamma$) reaction rate and shows agreement to NON-SMOKER predictions as well to \cite{Kim_2022} within the uncertainty limits. In contrast, there is little to no experimental information on the $^{59}$Cu($p$,$\alpha$)$^{56}$Ni reaction rate and this rate remained highly uncertain. In fact, this recent new measurement of ($p$,$\gamma$) emphasizes the need for new $^{59}$Cu($p$,$\alpha$)$^{56}$Ni studies \citep{OShea25}. Therefore, to understand the NiCu cycle strength and its impact on model-observation comparisons for XRBs, it of utmost importance to have an experimental determination of this rate with stringent constraints on uncertainty limits. Direct measurements are extremely challenging as $^{59}$Cu is short-lived.\\
In this work, we report on the direct measurement of $^{59}$Cu($p$,$\alpha$)$^{56}$Ni in inverse kinematics using $^{59}$Cu radioactive ion beam and a pure H$_2$ target. A previous direct measurement of $^{59}$Cu($p$,$\alpha$)$^{56}$Ni was reported, albeit at a higher center-of-mass energy (E$_{c.m.} $) of 6 MeV  \citep{Randhawa_59Cu}. Using the same experimental set-up, here we report direct measurements down to E$_{c.m.}=$ 3.6 MeV. Subsequent sections explain the experimental details, new results and their impact on model light curves and the NiCu cycle in  XRB bursts.

\begin{figure}
    \centering
    \includegraphics[width=\linewidth]{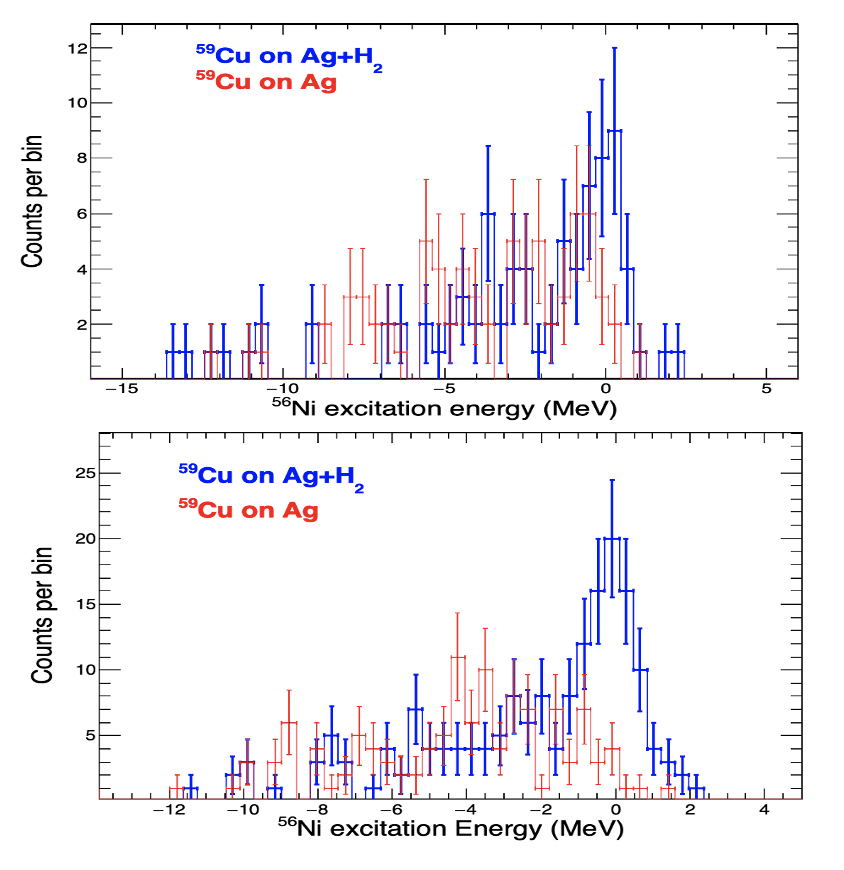}
    \caption{The excitation energy spectrum of $^{56}$Ni with (blue) and without (red) H$_{\rm{{2}}}$ target at E$_{c.m.}$ 4.0$\pm$0.4 MeV (upper) and 4.68$\pm$0.25 MeV (lower panel), respectively.}
    \label{fig2}
\end{figure}

\begin{figure}
    \centering
    \includegraphics[width=\linewidth]{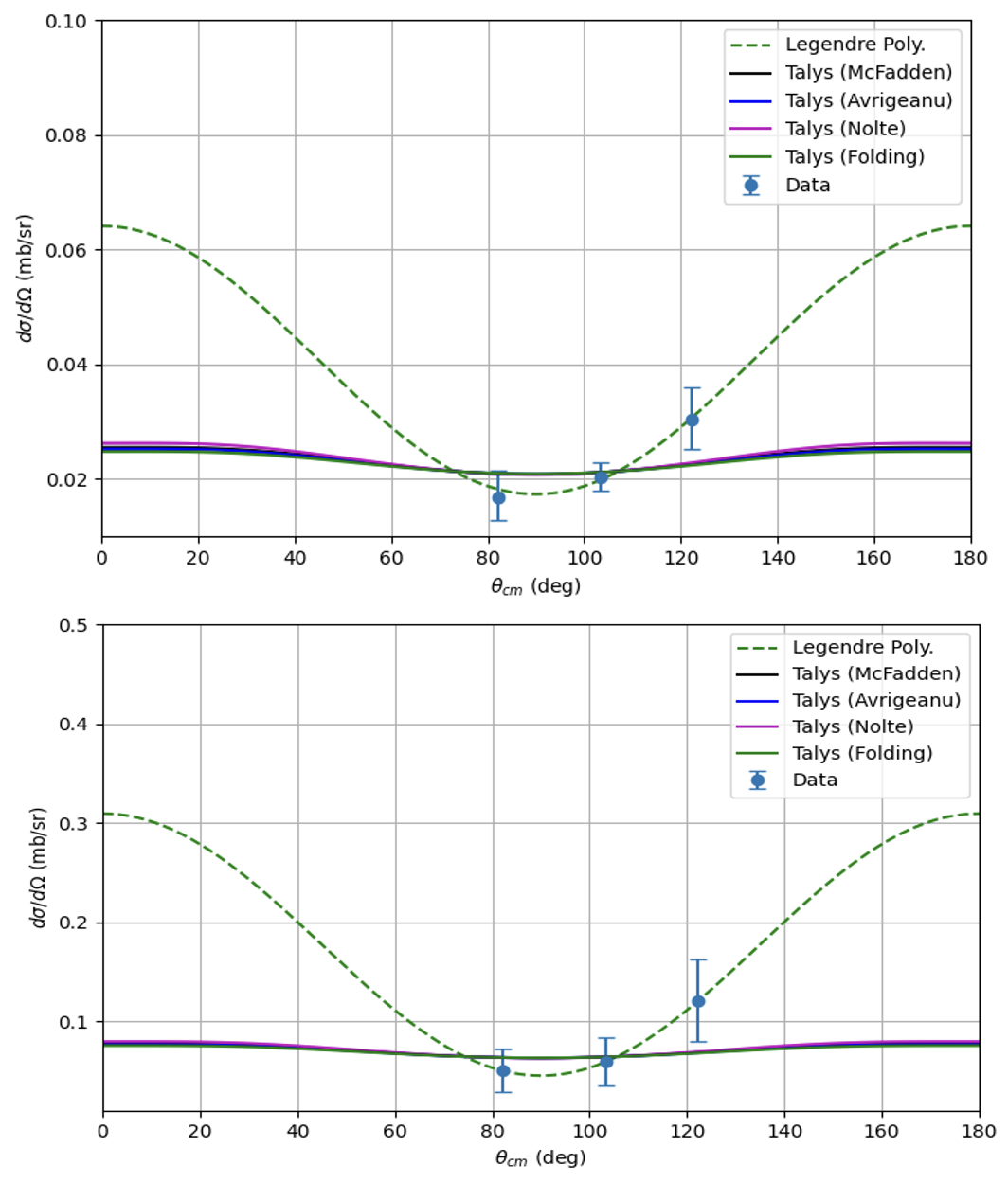}
    \caption{Experimentally determined angular distributions at E$_{c.m.}$ 4.0 MeV (upper panel), and at 4.68 MeV (lower panel). TALYS calculations (for different $\alpha$-OMPs) and Legendre polynomial fits to data are shown at both energies. }
    \label{fig3}
\end{figure}
\begin{figure}
    \centering
    \includegraphics[width=\linewidth]{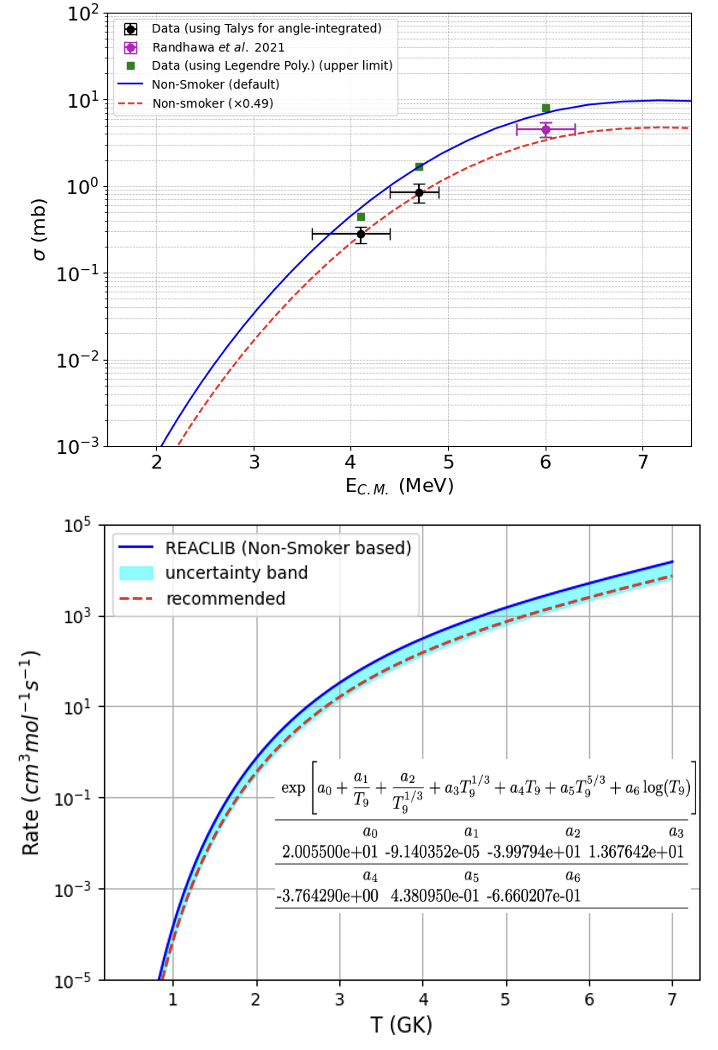}
    \caption{Top Panel: Experimental cross sections at the weighted center-of-mass energies are shown in comparison to default NON-SMOKER cross sections (blue curve). Lower Panel: New recommended reaction rate  and associated uncertainty band is shown in comparison to REACLIB rate which is based on Non-Smoker calculations. Lower panel inset shows REACLIB rate coefficients for
recommended rate.}
    \label{fig4}
\end{figure}
\section{Experiment} 
The experiment was performed using the IRIS facility at ISACII at TRIUMF. For a full schematic of IRIS facility, please refer to Figure 1 in \cite{Randhawa_59Cu}. A radioactive beam of $^{59}$Cu was produced and accelerated using the superconducting LINAC to 6.7$A$ MeV. The beam then passed through an ionization chamber, filled with isobutane gas, operated at two pressures, 19.5 Torr and 10 Torr, leading to two different beam energies at the center of the H$_{\rm{2}}$ target. The energy loss of the beam measured in this ionization chamber provided an event-by-event identification of the $^{59}$Cu incident beam and its contaminant $^{59}$Co throughout the experiment. Ionization chamber operation at these two pressures resulted in laboratory energies of $E_{\mathrm{lab}} = 238.7~\text{and}~278.5~\mathrm{MeV}$, corresponding to center-of-mass energies of $E_{\mathrm{cm}} = 4.0\pm0.4~\text{MeV and}~4.68\pm0.25~\mathrm{MeV}$ at the center of the H$_{2}$ target, respectively. The average beam intensities were $\sim$ $7.4\times10^{3}$~pps and $1.09\times10^{4}$~pps for the 4.0$\pm$0.4~MeV and 4.68$\pm$0.25~MeV runs, respectively. The beam impinged on a thin windowless solid hydrogen (H$_{2}$) reaction target with a 4.5~$\mu$m thick Ag foil backing facing upstream of the H$_{2}$ layer. The target cell with the foil was cooled to $\sim$4 K, and hydrogen gas is sprayed on the cooled Ag foil, which made a frozen/solid H$_{2}$ target. The solid H$_{\rm{2}}$ target has been successfully used in various experiments \citep{Junki, Randhawa_59Cu}. The energy of the beam elastically scattered from the Ag foil was measured using a downstream silicon strip detector detector, providing continuous measurement of the target thickness during the experiment. The average target thicknesses were 60.6~$\mu$m and 37.0~$\mu$m for the 4.0~MeV and 4.68~MeV data collection, respectively. At these energies, the open reaction channels were ($p,p$), ($p,p'$), and ($p$,$\alpha$). Protons and $\alpha$-particles from reactions were detected using annular arrays of 100 $\mu$m thick single-sided silicon strip detectors followed by a layer of 12 mm thick CsI(Tl) detectors, placed 15.5 cm downstream of the target. This detector combination served as an energy-loss and total energy (E) telescope for identifying the $p$ and $\alpha$-particle recoils after the target. The $\Delta$E-E correlation provided clear identification of the protons and $\alpha$-particles (Figure~\ref{fig1}). Any background contributions from reactions on the Ag foil were subtracted by using data without an the H$_{\rm{2}}$ target i.e., for beam impinging on the Ag foil alone. The detector telescope covered scattering angles of $\theta_{lab}$ = 18\textdegree{}–40.0\textdegree{}. 

\section{Analysis and Results}
The excitation energy spectra of $^{56}$Ni, shown in Fig.~\ref{fig2}, were reconstructed using the missing mass technique with the energy and scattering angle of the $\alpha$-particles, identified using the $\Delta$E-E (Fig.~\ref{fig1}) silicon-CsI(Tl) telescope. The peak around zero excitation energy in the spectrum is the ground state of $^{56}$Ni. The energy of the first excited state in $^{56}$Ni is 2.7 MeV and should be easily resolved from the ground state in the current experiment. However, the first excited state was not seen in the current measurement, which is in-line with the statistical model predictions that $^{59}$Cu(p,$\alpha$)$^{56}$Ni(g.s.) dominates the total cross section. A  main source of background is $\alpha$-particles originating from the reactions on the Ag foil. This background was measured by collecting data without the H$_{2}$ target and is shown with a red dashed-dotted histogram in Fig.~\ref{fig2}. Our angular coverage was $\sim$71\textdegree{}–131\textdegree{}, in the center-of-mass frame, for $\alpha$-particles identified through $\Delta$E-E telescope. Therefore, to get angle-integrated cross sections, these data were used to extract the angular distributions for further analysis. Extracted angular distributions at the two different center-of-mass energies are shown in Figure~\ref{fig3}. To get the angle integrated total cross sections we calculated angular distributions using the code TALYS with different $\alpha$-OMPs, namely the McFadden and Satchler, Folding, Nolte, and Avrigeanu potentials \citep{talys}. The calculated curves were normalized to best fit the data through chi-square minimization (Fig.~\ref{fig3}). The best fit angular distribution curves were used to obtain the angle integrated cross sections. To report the final cross sections, we used angle-integrated cross sections obtained using scaled angular distributions from McFadden $\alpha$-OMPs, and the difference between the angle-integrated cross sections using different $\alpha$-OMPs was taken as a part of the systematic uncertainty analysis. The total cross sections extracted at $E_{\mathrm{cm}} = 4.0\pm 0.4 ~\text{and}~4.68\pm0.25 ~\mathrm{MeV}$ were 0.28$\pm$0.06 mb, and 0.85$\pm$ 0.21 mb, respectively. In addition to the TALYS calculations to obtain angle integrated cross sections, we also used Legendre polynomial fits  
(which are frequently used to obtain angle-integrated cross sections \citep{Ahn,19F_pa,Voinov}): 
$\frac{dY}{d\Omega} = \sum_{l=0,\,\text{even}}^{l_{\text{max}}} a_l P_l(\cos\theta)$, 
where $P_{l}(\cos\theta)$ are the Legendre polynomials of order $l$, $l_{\text{max}} = 2$ in this work, and $a_{l}$ are the angular distribution coefficients that were treated as free parameters in the fitting. As angular distributions are expected to be symmetric about 90 deg., only even order coefficients are possible leading to only 2 free parameters. This fitting function accounted for more anisotropy compared to Talys calculations and could describe the data equally well compared to Talys predictions. It should be noted that for the p+$^{59}$Cu system, level densities are expected to be high enough that statistical models are appropriate to describe the data. Therefore, we have used the TALYS calculations for angular distributions to obtain the recommended rate and the one obtained from Legendre Polynomials was only used as an upper limit. Angle-integrated cross sections obtained using Legendre Polynomials are shown in Fig.~\ref{fig4} (green dots), which was taken as a cross section upper limit (hence no associated uncertainty is shown) in this work and agrees with the theoretical NON-SMOKER predictions. To consider the change in cross section over the target thickness, we have considered the weighted energies, defined as $\int \sigma(E)E dE/\int \sigma(E)dE$ (where the energy dependence of HF based NON-SMOKER cross sections was used). Weighted center-of-mass energies were 4.1 MeV and 4.71 MeV for 4.0 MeV and 4.68 MeV, respectively. Comparison of new cross sections to NON-SMOKER based cross sections are shown in Figure~\ref{fig4} (upper panel), where data points are shown at the weighted energies. Scaled NON-SMOKER cross sections (best fit to data) are also shown in comparison. The scaling factor required to achieve a best fit was 0.49 (dotted red line) when weighted energies are considered. The scaled NON-SMOKER cross-section was used to obtain the reaction rate and is shown along with associated uncertainty in Figure ~\ref{fig4} (lower panel) in comparison to the REACLIB rate. The reaction rate uncertainty band (cyan color) accounts for the upper limit corresponding to angle integrated cross sections from the Legendre polynomial and the lower limit of the scaled cross section sets the lower part. The recommended rate is shown as a dashed red line. Figure~\ref{fig4} (lower panel inset) shows the best-fit parameters for the new rate as a fit to standard seven parameter REACLIB format. Our new recommended rate is a factor of 2 lower compared to the NON-SMOKER prediction but the upper limit does not rule out this prediction. The reported new reaction rate with its associated uncertainty (factor $\sim$2) brings stringent constraints on the $^{59}$Cu(p,$\alpha$) reaction rate, compared to the previously used uncertainty limits of factor 10$^{4}$ (100 up and down).  

\begin{figure}
    \centering
    \includegraphics[width=\linewidth]{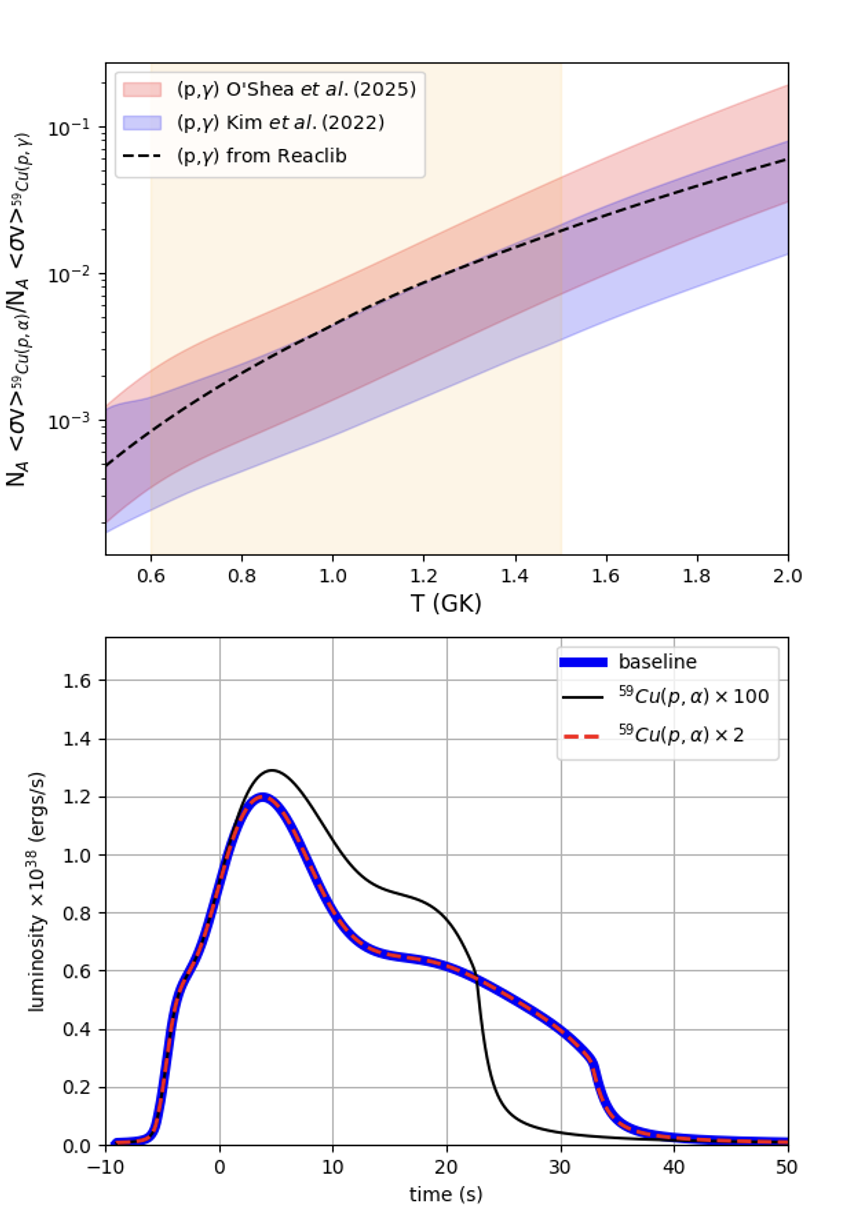}
    \caption{Top Panel: Ratio of  $^{59}$Cu(p,$\alpha$)$^{56}$Ni (recommended rate, this work) to $^{59}$Cu(p,$\gamma$)$^{60}$Zn from two different works, depicting NiCu cycle strength in XRBs. Temperatures of interest for XRBs is highlighted by the yellow band. Lower panel: Impact of reaction rate variation on one-zone X-ray burst light curves, with ignition conditions curated for X-ray binary $\tt{GS 1826-24}$. New rate, considering upper-limit, is factor 2 uncertain, and this change in reaction rate does not impact the light curve shape.}
    \label{fig5}
\end{figure}
\section{Implications for X-ray bursts}
\subsection{Impact on Ni-Cu cycle}
The strength of the NiCu cycle can be characterized by the ratio of the $^{59}$Cu($p$,$\alpha$)$^{56}$Ni rate to the $^{59}$Cu($p$,$\gamma$)$^{60}$Zn rate \citep{Kim_2022}.  Our new measurement rules out any enhancement in the reaction rate beyond the NON-SMOKER based rate currently used in the reaction rate libraries. The previously determined $^{59}$Cu($p$,$\gamma$)$^{60}$Zn  reaction rate from the existing $p$+$^{59}$Cu resonance data \citep{Kim_2022}, provided the upper and lower limit to reaction rate and  agrees with the statistical model predictions within uncertainty limits. For the recent indirect measurement of $^{59}$Cu($d$,n$\gamma$) \citep{OShea25}, which extracted the proton spectroscopic factors, we assumed a factor of 5 uncertainty (as factor 5 uncertainty is not ruled out by the authors) in the recommended rate calculated using Table 2 of \citep{OShea25}. Within these uncertainty limits their extracted ($p$,$\gamma$) rate agrees with NON-SMOKER predictions and as well as  with the reaction rate from \cite{Kim_2022}. Fig.~\ref{fig5}(upper panel) shows the ratio of  $^{59}$Cu(p,$\alpha$)$^{56}$Ni  using the current recommended rate to $^{59}$Cu(p,$\gamma$)  rate from \cite{OShea25} and from \cite{Kim_2022}  and their associated uncertainty, which depicts the strength of NiCu cycle. The ratio even at the higher end of temperature relevant for XRBs (at 1.5 GK) shows that the recycling of material in NiCu cycle is negligible, $\leq$5\%. The current measurement thus rules out the possibility of a strong NiCu cycle in X-ray bursts and in fact, shows that the NiCu recycling is too weak or negligible.  

\subsection{Impact on XRB light curves}
To study the impact of the current measurement on X-ray burst observables, we employ a one-zone model ($\tt{onezonerp}$ by \cite{cumming_onezonerp}) and adopt the new reaction rate (recommended rate in Fig.~\ref{fig4}) as the nominal value. The ignition conditions for the one-zone model were used from Ref.\citep{Cyburt_2016}, where these were calibrated using the KEPLER 1-D multi-zone X-ray burst model, to reproduce the light curve and final composition for the clocked-burster GS 1826$-$24 . The one-zone model then follows a single burst considering nuclear energy generation and cooling. As the nominal rate is factor of $\sim$2 lower compared to NON-SMOKER calculations, we varied the ($p$,$\alpha$) reaction by factor of 2 upwards. For comparison, the impact of varying the $^{59}$Cu($p$,$\alpha$) by a factor 100 up is also shown, as was used in the sensitivity study of Cyburt $et$ $al.$\citep{Cyburt_2016}.  Within the new variation limit of a factor 2 up, the $^{59}$Cu($p$,$\alpha$) reaction does not show any impact on XRB light curve, as shown in Fig.~\ref{fig5} (lower panel). This measurement, thus removes one of the most important nuclear physics uncertainties in the XRB models, hence facilitating the future model-observation comparison.
\vspace{1pt}
\section{Summary}
To summarize, nuclear reactions in NiCu cycle in the X-ray bursts, featuring competition between $^{59}$Cu(p,$\alpha$) and  $^{59}$Cu(p,$\gamma$) are some of the most important nuclear physics uncertainties in the XRB models, hence a hindrance in the model-observation comparison. We report direct measurement of $^{59}$Cu($p$,$\alpha$) using a radioactive $^{59}$Cu beam at TRIUMF. The new measurement brings stringent constraints on this reaction rate and in conjuction with recent ($p$,$\gamma$) reaction rate precludes a strong NiCu cycle in XRBs. The new rate, when varied within the current uncertainty, does not have any impact of the XRB light curve. This measurement removes an important nuclear physics uncertainty in the XRB models, and thus facilitates  future model versus observation comparisons, necessary to extract the astrophysical conditions and neutron star compactness.


\begin{acknowledgments}
We gratefully acknowledge the support of the beam delivery team at TRIUMF. Support from NSERC, CFI, and Research Nova Scotia is gratefully acknowledged. JSR acknowledges the support from Office of Research and Economic Development, and College of Arts and Sciences at Mississippi State University. TRIUMF receives funding via a contribution through the National Research Council Canada. This work was also supported by the National Science Foundation under grant no. NSF-PHY 2011890.
\end{acknowledgments}


\bibliography{bibliography}{}
\bibliographystyle{aasjournalv7}



\end{document}